\input epsf
\documentclass[aps,prb,twocolumn]{revtex4}
\def\tens#1{\underline{\underline{#1}}}
\begin{document}

\title{Ab-initio study of the vibrational properties of Mg(AlH$_4$)$_2$}
\author{E. Span\`o  and M. Bernasconi}

\affiliation{Consorzio Corimav, Dipartimento di Scienza dei Materiali and Istituto Nazionale di Fisica
per la Materia,  Universit\`a di  Milano-Bicocca, Via Cozzi 53, I-20125, Milano, Italy}

\begin{abstract}
 Based  on Density Functional Theory and Density Functional 
 Perturbation Theory  we have studied the
 thermodynamical and vibrational properties of
 Mg(AlH$_4$)$_2$.
 The crystal structure
 recently  proposed  on the basis of x-ray powder diffraction data has been confirmed theoretically
 by the comparison of  
 the  experimental and theoretical IR and Raman spectra.
 The main discrepancy regards the position of the hydrogen atoms which makes the theoretical AlH$_4$ tetrahedra
 more symmetric than the experimental ones.
 The calculated thermodynamical decomposition temperature  
 is  also in good agreement with experimental result.
\end{abstract}

\pacs{???}

\maketitle

\section{Introduction}

In the search for suitable materials for reversible hydrogen storage the class of metal alanates 
   has attracted much attention in recent years due to their
high hydrogen content, but primarly because of the possibility to accelerate the kinetics of hydrogen uptake
by doping \cite{bodga1,bogda2}. 
Sodium alanate, the prototypical material  of this class, has long been known. Although its thermodynamics 
 decomposition temperature with hydrogen release is relatively low, the 
kinetics of the dehydrogenation/rehydrogenation reactions is very slow. 
The interest in this material for hydrogen storage increased dramatically in 1997 when Bogdanovic {\sl et al.}
\cite{bodga1} demonstrated that reversible storage can be achieved at moderate temperature and hydrogen partial 
pressure by the addition of catalysts, most notably titanium.
This discovery boosted an intense activity on the study of other members of the alkali metal alanates as well
\cite{LiAl,KAl}. 

More recently, magnesium alanate Mg(AlH$_4$)$_2$, as a representative of the  alkali earth metal alanates, has also
been considered as a possible materials for hydrogen storage \cite{ficth0,ficth1,ficth2,ficth3}.
Mg(AlH$_4$)$_2$ decomposes readly in the temperature range 110 $^o$C - 200 $^o$C according to the reaction

\begin{equation}
 Mg(AlH_4)_2 \rightarrow MgH_2 + 2 Al + 3 H_2 
\label{reaction}
\end{equation}

which would correspond to a maximum reversible hydrogen content of 7 wt$\%$.
Although the reversibility of reaction \ref{reaction} has not been demonstrated yet, the easy
of the decomposition reaction even in the absence of doping, suggests that, in analogy with 
NaAlH$_4$, the insertion of suitable catalyst might enhance the kinetics of 
hydrogen uptake.
The structure of  Mg(AlH$_4$)$_2$ has been recently resolved from X-ray powder diffraction pattern
aided by quantum-chemical calculations on cluster models which allowed to tentatively assign also the position of hydrogen
atoms, not detectable by X-ray diffraction \cite{ficth3}. 

In this work, we present an ab-initio study of the thermodynamical and vibrational properties of
Mg(AlH$_4$)$_2$ aiming at providing a better estimate of the decomposition temperature from calculations on
periodic models and at confirming the structure inferred experimentally from the comparison of  theoretical
and experimental IR and Raman spectra.

\section{Computational details}

Calculations are performed within the framework of Density Functional Theory (DFT)  with
a gradient corrected  exchange and correlation energy functional  \cite{blyp},
as implemented in the codes PWSCF and PHONONS \cite{pwscf}.
Norm conserving pseudopotentials \cite{TM} and plane wave expansion of 
Kohn-Sham (KS) orbitals up to a kinetic cutoff of 40 Ry have been used.
Non linear core 
corrections are included in the pseudopotential of magnesium \cite{nlcc}.
Brillouin Zone (BZ) intergration has been performed over  Monkhorst-Pack (MP)
\cite{MP}  6x6x6, 4x4x4 and 16x16x16 meshes for Mg(AlH$_4$)$_2$, MgH$_2$ and metallic Al,
respectively. 
Hermite-Gaussian smearing \cite{meth} of order one with a linewidth of 0.01 Ry has been used in the reference calculations 
on metallic Al.
Equilibrium geometries have been obtained by 
optimizing the internal and lattice structural parameters at several volumes and fitting the energy
versus volume data with a Murnaghan function \cite{murna}.
Residual anisotropy in the stress tensor at the optimized lattice parameters at each volume is below 0.6 kbar.
Infrared and Raman spectra are obtained from effective charges, dielectric susceptibilities and 
phonons at the $\Gamma$ point within density functional perturbation theory \cite{dfpt}.
Since Mg(AlH$_4$)$_2$ is an uniaxial crystal, the infrared absorption spectrum depends on
the polarization of the trasmitted electromagnetic wave with respect to the optical axis.
Relevant formula for the calculation of the IR and Raman spectra for a polycristalline sample to be compared with
experimental data are given in section II.

\section{Results}

\subsection{Structure and energetics}

The structure of Mg(AlH$_4$)$_2$ has been recently assigned 
to a CdI$_2$-like layered crystal 
from X-ray powder diffraction data \cite{ficth3}.
Rietveld refinement assigned the space group $P{\bar 3}m1$ ($D_{3d}^3$)
and lattice parameter $a$=$b$=5.199 $\rm\AA$ and $c$=5.858 $\rm\AA$.
The position of the four independent atoms in the unit cell are given in Table I.
In the AlH$_4$ tetrahedron there are two Al-H bond lengths: the length of the Al-H1 bond  with non-bridging H1 hydrogen 
aligned with the $c$ axis and the length of Al-H2 bond with the H2 atom bridging between Al and Mg.

\begin{table}
\caption{Experimental  and theoretical (in parenthesis) positions (in crystal units) of the four independent atoms
at the experimental equilibrium lattice parameters (space group $P{\bar 3}m1$, $a$=5.199 $\rm\AA$, 
$c$=5.858 $\rm\AA$). The Wyckoff notation is used. If not reported the theoretical positions coincide with
the experimental ones by symmetry. H1 and H2 are non-bridging and  bridging hydrogen atoms, respectively.}
\begin{ruledtabular}
\begin{center}
\begin{tabular}{lccc}

Mg(1a)  & 0 & 0& 0 \\
Al(2d)  & 1/3 & 2/3 & 0.7 (0.7053) \\
H1(2d)  & 1/3 & 2/3 & 0.45 (0.4349) \\
H2(6i)  & 0.16 (0.1678) & -0.16 (-0.1678) & 0.81 (0.8111) \\
   \end{tabular}
   \end{center}
   \end{ruledtabular}
   \label{position}
   \end{table} 

\begin{figure}[!h]
\centerline{\epsfysize= 5. truecm\epsffile{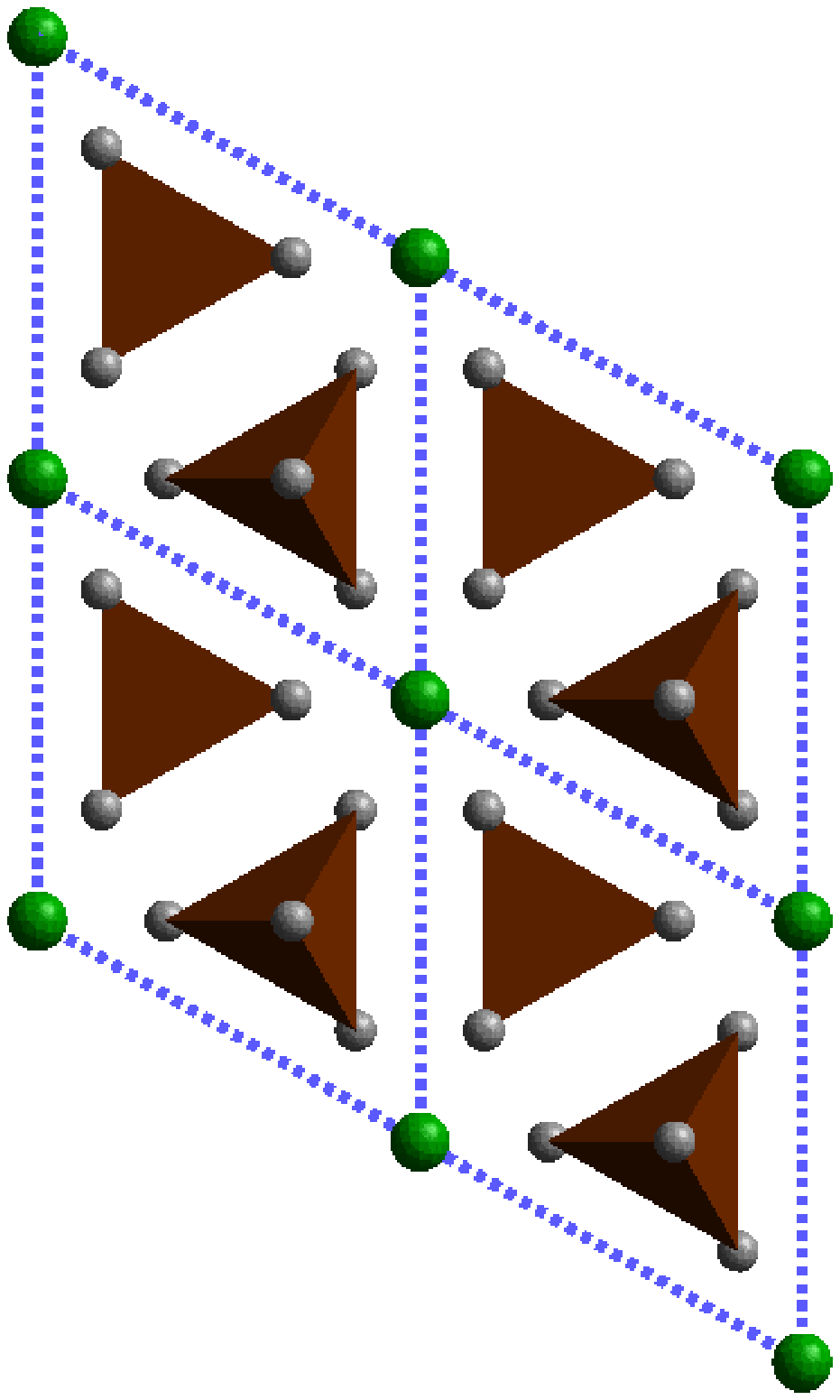}}
\centerline{\epsfysize= 5. truecm\epsffile{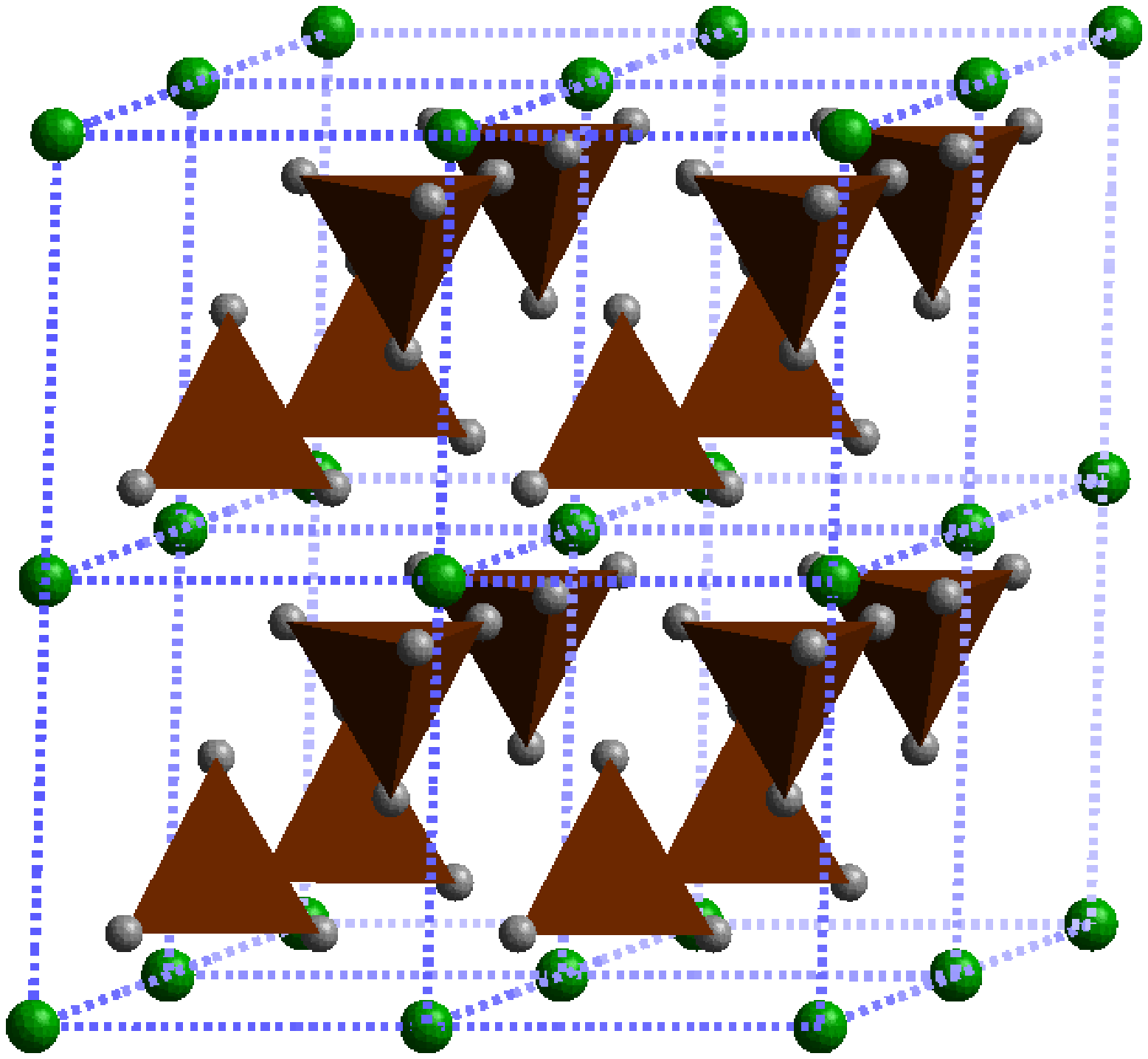}}
\caption{(Color online) Trigonal structure of Mg(AlH$_4$)$_2$. The unit cell containing a single
formula unit is drawn.}
 \label{structure}
\end{figure}

A picture of the crystal structure is given in Fig. \ref{structure}. It can be seen as a stacking along the $c$ axis
of AlH$_4$-Mg-AlH$_4$ neutral trilayers.

The theoretical equilibrium lattice parameters obtained from the Murnaghan equation of state are
$a$=5.229 $\rm\AA$ (exp. 5.199) and $c$= 6.238  $\rm\AA$ (exp. 5.858).
 In the geometry optimization we have also relaxed the constrained of 
$P{\bar 3}m1$ symmetry, but we have always recovered the experimental space group.
The misfit in the $c$ axis ( 6 $\%$) is  larger than usual in DFT-based calculations.
As already pointed out in previous cluster calculations on Mg(AlH$_4$)$_2$ \cite{ficth3}
the trilayers stacked along the $c$ axis are neutral and not chemically bonded each other.
Thus van der Waals interactions have been proposed to  play an important role in interlayer cohesion.
Since van der Waals interations are not present in current approximations to the exchange and correlation
energy functional, the interlayer bonding is underestimated which implies an expansion of the $c$ axis with
respect to the experimental data. The energy difference between the relaxed configurations at the
experimental and theoretical equilibrium lattice parameters is nevertheless small ( 38 meV per formula unit).
The theoretical atomic positions at the experimental lattice parameters are given in Table I.
The theoretical geometry of the AlH$_4$ tetrahedron is closer to an ideal tetrahedron than the experimental geometry.
In fact 
the theoretical length of the two  Al-H bonds  are Al-H1= 1.584 $\rm\AA$ (exp. 1.46) and Al-H2= 1.614 $\rm\AA$ (exp. 1.69).
The Al-H bond lengths do not change sizably (within 0.01 $\rm\AA$) by changing the lattice parameters from the
experimental values to the theoretical equilibrium values. 

\begin{figure}[!h]
\centerline{{ \epsfysize= 5. truecm\epsffile{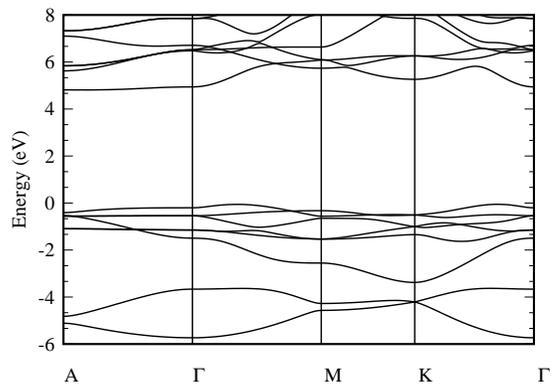}}}
\caption{Electronic band structure along the high symmetry direction of the Brillouin Zone.
High symmetry points are labeled following Zak (Ref. \cite{zak}).
The zero of energy is the top of the valence bands.}
 \label{bands}
\end{figure}

The electronic band structure along the high symmetry directions of the 
Irreducible Brillouin Zone is reported in Fig. \ref{bands}.
The analysis of the projection of the KS states on the atomic orbitals at the $\Gamma$ point shows that
the top of the valence band is mainly formed by  3p states of aluminum and 1s state of hydrogen while the bottom of the conduction band is
mainly formed by 3s states of magnesium. The band gap is thus a charge transfer excitation. 
\vspace{1.cm}

By neglecting at first the vibrational contribution to the entropy of the solids, the thermodynamical decomposition
temperature $T_D$ of reaction \ref{reaction} can be estimated as

\begin{equation}
T_D = \frac{\Delta E + P\Delta V}{3 (S_{{\rm H}_2}-R)} 
\label{TD}
\end{equation}

where $\Delta V$ is the difference in volume of the solid reactants and products of reaction \ref{reaction}, $P$ is
the pressure,
$\Delta E$ is the energy difference of reactants and products in \ref{reaction}, $S_{{\rm H}_2}$ is the entropy per mole of gaseus
 H$_2$ and the gas constant $R$ in the denominator comes from the PV term of gaseus H$_2$.
$\Delta V$  and $\Delta E$ are obtained from the ab-initio calculations.
For MgH$_2$ the theoretical equilibrium structural parameters are $a$= 4.470  $\rm\AA$ (exp. 4.501 
\cite{MgH2}), $c$= 2.943  $\rm\AA$ (exp. 3.010 \cite{MgH2}) and the 4$f$ position (in Wyckoff notation of the
$P4_2/mnm$ space group)  of the independent H atom in the unit cell is (0.3046,0.3046,0)
(exp. (0.3040,0.3040,0) \cite{MgH2}).
The theoretical equilibrium lattice parameter of metallic Al is $a$=  4.065  $\rm\AA$ (exp. 4.050  $\rm\AA$  \cite{ash}).
From the total energies at equilibrium we obtain $\Delta E$= 144 kJ/mol (48 kJ/mol per H$_2$ molecule) which becomes 
 $\Delta E$= 139.5  kJ/mol  (46.5 kJ/mol per H$_2$ molecule) 
 by including the zero point energy of the optical phonons  (at the $\Gamma$ point only)
 and of the acoustic bands within a Debye model \cite{notadebye}.
 This result is in good agreement with previous estimate 
from the extrapolation of cluster calculations (123 kJ/mol \cite{ficth3}, 41 kJ/mol per H$_2$ molecule).
The $P\Delta V$ in Eq. \ref{TD} is negligible (4.4 J/mol).
By assuming $S_{{\rm H}_2}$= 130 J/mol K$^{-1}$ at 300 K \cite{SH} and atmospheric pressure, we obtain
$T_D$= 111 $^o$C. 
The predicted  T$_D$ shifts by 2 $^o$C
by including the temperature dependence of $S_{{\rm H}_2}$ as given by the Sackur-Tetrode expression
\cite{atkins}, the vibrational contribution to the free energy of the solids and the translational and rotational energy of
gaseous H$_2$ \cite{notadebye}. 
The theoretical $T_D$ is comparable with the experimental decomposition temperature which falls in the range  
110 $^o$C - 200 $^o$C which implies that kinetics effect do not affect  dramatically   the decomposition reaction of
Mg(AlH$_4$)$_2$.

\subsection{Vibrational properties}

Phonons  at the $\Gamma$-point can be classified according to the irreducible representations of
the $D_{3d}$ point group of Mg(AlH$_4$)$_2$ as  $\Gamma$= 4 A$_{1g}$ + 2 A$_{2g}$ + 5 E$_{g}$ + 5 A$_{2u}$ + 6 E$_{u}$.
 A$_{1g}$ and  E$_{g}$  modes are Raman active while  A$_{2u}$ and  E$_{u}$ modes are IR active. One A$_{2u}$ and  one E$_{u}$ mode
 are uniform translational modes.
The IR active modes display a dipole moment which couple to the inner macroscopic 
longitudinal field which shifts the LO phonon frequencies via the non-analytic contribution to
the dynamical matrix \cite{dfpt}

\begin{equation}  
D^{NA}_{\alpha,\beta}(\kappa,\kappa')=\frac{4 \pi}{V_o} \frac{Z_{\alpha,\alpha'}(\kappa)q_{\alpha'}
Z_{\beta,\beta'}(\kappa')q_{\beta'} }
{{\bf q} \cdot \tens{\varepsilon}^{\infty}  \cdot {\bf q}},
\label{macro}
\end{equation}

where $\tens{Z}$ and $\tens{\varepsilon}^{\infty}$ are the effective charges and electronic dielectric tensors, $V_o$ is the
unit cell volume and 
{\bf q} is the phononic wavevector.
The effective charge tensor for the three independent atoms in the unit cell (cfr. Table \ref{position}) are 

\begin{equation}
Z(Mg)=\left(\begin{array}{ccc}
         2.180    &    &     \\
	                         & 2.180  &     \\ 
                           &     & 1.974 \\
            \end{array}  \right), \ \ \
                 \label{ZM} 
                     \end{equation}

\begin{equation}
Z(Al)=\left(\begin{array}{ccc}
    1.934        &    &     \\
	                         & 1.934  &     \\ 
                           &     &  1.678\\
            \end{array}  \right), \ \ \
                 \label{ZA} 
                     \end{equation}

\begin{equation}
Z(H1)=\left(\begin{array}{ccc}
         -0.627   &    &     \\
	                         & -0.627  &     \\ 
                           &     &  -0.641 \\
            \end{array}  \right), \ \ \
                 \label{H1} 
                     \end{equation}

\begin{equation}
Z(H2)=\left(\begin{array}{ccc}
      -0.564      &    &     \\
	                         & -1.029  & 0.248    \\ 
                           &  0.289   &  -0.672 \\
            \end{array}  \right), \ \ \
                 \label{H2} 
                     \end{equation}

The electronic dielectric tensor is

\begin{equation}
\tens{\varepsilon}^{\infty}=\left(\begin{array}{ccc}
          2.894   &    &  \\
	               &  2.894  &   \\
		                &    & 2.781  \\
			              \end{array}  \right)\ \ \ ,
				      \label{epsi81}
				      \end{equation}

The calculated phonon frequencies at the $\Gamma$ point, neglecting the contribution of the 
longitudinal macroscopic field (Eq. \ref{macro}), are reported in Table II for the equilibrium geometry
at the experimental lattice parameters. The phonon frequencies at the theoretical lattice parameters differ at most
by 10 cm$^{-1}$ with respect to those reported in Table II.

\begin{table}
\caption{Theoretical phonon frequencies at the $\Gamma$ point, 
oscillator strengths ($f_j$ in Eq.
\ref{epsiperp}) of IR $u$-modes and 
coefficients of the Raman tensor of the
Raman active mode, $a$, $b$ for A$_{1g}$ and $c$, $d$ for E$_g$ modes (in unit of 10$^{-3}$ $V_o$=
0.2743 $\rm\AA^3$, see section III B.2). 
The contribution of the inner longitudinal macroscopic field is not included (LO-TO splitting) (see Fig. \ref{disp} for the
displacement patterns).}
\begin{ruledtabular}
 \begin{center}
   \begin{tabular}{lccccc}
   Modes  & Energy  (cm$^{-1}$) & f$_j$  & a (c) & b (d) &                    \\

    E$_{g}$ (1)   &          87 &  & 0.976 & 0.578\\
    A$_{2g}$ (1)   &         169 &  \\
    A$_{1g}$ (1)   &         232 &  & 2.337 & 1.210\\
    E$_{u}$ (1)  &         282 &  3.162\\
    E$_{g}$ (2)  &         298 & \\
    A$_{2u}$ (1) &         302 & 0.103 \\
    A$_{2g}$ (2)  &         355 & \\
    E$_{u}$ (2) &         360 &  0.758\\
    E$_{u}$ (3) &         620 & 3.896 \\
    A$_{2u}$ (2)  &         663 &  0.784 \\
    E$_{u}$ (4) &         716 & 0.064\\
    E$_{g}$ (3) &         742 & & -0.457  & 3.653  \\
    E$_{g}$ (4)  &         758 & &   2.746 & 0.023\\
    A$_{1g}$ (2)  &         812 & &  3.970  & 0.681  \\
    A$_{1g}$ (3)  &        1845 &  & 8.768  & 17.540 \\
    A$_{2u}$ (3)  &        1850 & 0.072 \\
    E$_{g}$ (5)  &        1852 & & 5.202    & 0.339 \\
    E$_{u}$ (5) &        1905 & 0.597\\
    A$_{2u}$ (4)  &        2013 &   0.040 \\
    A$_{1g}$ (4) &        2077 &    & 12.270  & 5.160 
     \end{tabular}                  
      \end{center}                              
\end{ruledtabular}
  \label{phonon}
   \end{table} 

\begin{figure}[!h]
\centerline{\epsfysize= 6 truecm\epsffile{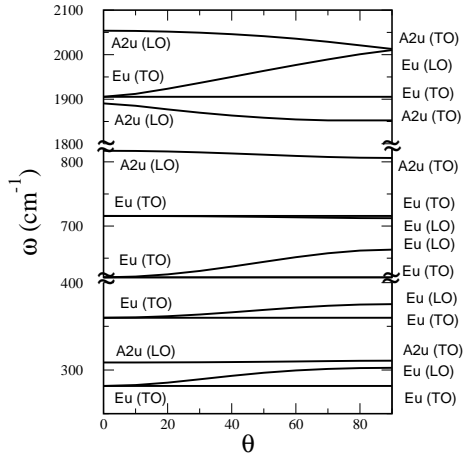}}
\caption{Angular dispersion of the $\Gamma$-point phonons. Only A$_{2u}$ and E$_{u}$ modes have angular dispersion.
$\theta$ is the angle formed by the phonon wavevector and the optical axis.}
 \label{adisp}
\end{figure}

For a uniaxial crystal like Mg(AlH$_4$)$_2$,  the macroscopic field
contribution to the dynamical matrix (Eq. \ref{macro})
introduces an angular dispersion of the phonons at the $\Gamma$ point, i.e. the limit of the phononic
 bands  
 $\omega({\bf q})$ for ${\bf q} \rightarrow 0$ depends on the angle $\theta$ formed by {\bf q} with the optical axis.
The angular dispersion of the A$_{2u}$ and E$_{u}$ modes due to the macroscopic field is reported in Fig.
\ref{adisp}.

\subsubsection{IR spectrum}

Tha absorption coefficient of an uniaxial crystal depends on the 
the polarization of the trasmitted light with respect to the optical axis.
The optical properties of the crystal can be obtained from the two dielectric functions
$\epsilon_{\perp}(\omega)$ and $\epsilon_{\parallel}(\omega)$ which represent the response of the crystal to
electromagnetic wave  with electric field perpendicular
({\bf E} $\perp$ {\bf c}) and parallel  ({\bf E} $\parallel$ {\bf c}) to the optical axis, respectively.
The trasmitted electromagnetic wave
at a generic wavevector ${\bf q}$ forming an angle $\theta$ with the
optical axis would split in an ordinary wave with electric field perpendicular to the optical axis 
 and in an extraordinary wave
with electric field lying in the plane formed by the optical axis and ${\bf q}$.
The dielectric function $\epsilon_{\perp}(\omega)$, independent on $\theta$, describes the response to the ordinary
waves while the dielectric function $\epsilon_{\theta}(\omega)$ for the extraordinary wave is $\theta$-dependent 
and is given by

\begin{equation}
\epsilon_{\theta}(\omega)=\frac{\epsilon_{\perp}(\omega)\epsilon_{\parallel}(\omega)} 
{\epsilon_{\perp}(\omega)sin^2\theta + \epsilon_{\parallel}(\omega)cos^2\theta}.
\label{extra}
\end{equation}

The extraordinary wave coincides with the ordinary wave for $\theta=0$, i.e. propagation along the optical axis.


$\epsilon_{\perp}(\omega)$ and $\epsilon_{\parallel}(\omega)$ can be obtained from ab-initio phonons, effective charges
and electronic dielectric tensor  as

\begin{eqnarray}
\epsilon_{\perp}(\omega) & = &\epsilon^{\infty}_{\perp} + \frac{4 \pi}{V_o}\sum_{j=1}^{\nu} | 
\sum_{\kappa =1}^N  \tens{Z} \cdot \frac{{\bf e}(j,\kappa)}{\sqrt{M_{\kappa}}}|^2 \frac{1}{\omega^2_j -\omega^2} \nonumber \\
& = & \epsilon^{\infty}_{\perp} + \sum_{j=1}^{\nu} \frac{f_j \omega^2_j}{\omega^2_j -\omega^2}
\label{epsiperp}
\end{eqnarray}

\begin{equation}
\epsilon_{\parallel}(\omega)=\epsilon^{\infty}_{\parallel} + \frac{4 \pi}{V_o}\sum_{j=1}^{\mu} | \\
\sum_{\kappa =1}^N  \tens{Z} \cdot \frac{{\bf e}(j,\kappa)}{\sqrt{M_{\kappa}}}|^2 \frac{1}{\omega^2_j -\omega^2},
\label{epsipar}
\end{equation}

where $\nu$ and $\mu$ are the number of  E$_u$ and A$_{2u}$ TO modes, respectively.
The phonons entering in Eq. \ref{epsiperp} (\ref{epsipar}) have wavevector {\bf q} parallel (perpendicular) to the
optical axis. The notation $\parallel$ and $\perp$ refers to the orientation 
with respect to the optical axis of the dipole moment of the phonon
which coincides with that of the  electric field of the trasmitted wave it couples to.
The sum over $\kappa$ run over the $N$ atoms in the unit cell with mass $M_{\kappa}$. 
 ${\bf e}(j,\kappa)$ and $\omega_j$
are the eigenstates and eigenvalues of the dynamical matrix at the $\Gamma$ point,
without the contribution of the macroscopic field which has no effect on the 
purely TO modes. 
The absorption coefficient for the ordinary wave is given by

\begin{eqnarray} 
 \alpha_{\perp}(\omega) &  = & \frac{\omega}{nc} {\rm Im} \epsilon_{\perp}(\omega+i\gamma, \gamma \rightarrow 0) \nonumber \\
                  &  =  &  \frac{2 \pi^2}{V_onc}\sum_{j=1}^{\nu} | 
\sum_{\kappa =1}^N  \tens{Z} \cdot \frac{{\bf e}(j,\kappa)}{\sqrt{M_{\kappa}}}|^2 \delta(\omega-\omega_j),
\label{absorb} 
\end{eqnarray}

$c$ is the velocity of light in vacuum and $n$ is the refractive index.
The absorption coefficient for the extraordinary waves $\alpha_{\parallel}(\omega)$ with {\bf E} $\parallel$ {\bf c} 
(and {\bf q} $\perp$ {\bf c}) is given by
the analougous expression by changing $\nu$ with $\mu$ in the sum over phonons in Eq. \ref{absorb}.
The $\delta$-functions in Eq. 12 are approximated by Lorenztian functions
as

\begin{equation}    
\delta(\omega-\omega_j)=\frac{4}{\pi}\frac{ \omega^2 \gamma}{(\omega^2-\omega_j^2)^2 + 4 \gamma^2\omega^2}.
\label{lorenz}
\end{equation}

\begin{figure}
\centerline{\epsfysize= 9. truecm\epsffile{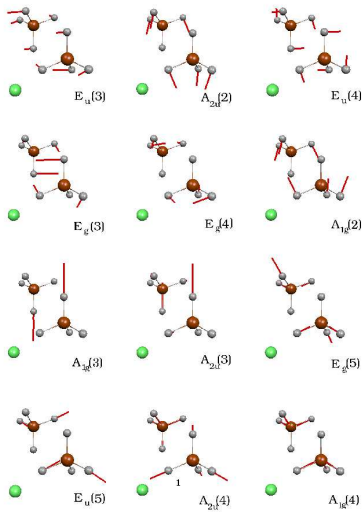}
}
\caption{(Color online) Displacement pattern of the IR and Raman active intra-tetrahedra modes. The lattice modes at lower frequencies are not
reported.}
 \label{disp}                                                                                                
 \end{figure}

\begin{figure}[!h]
\centerline{\epsfysize= 5. truecm\epsffile{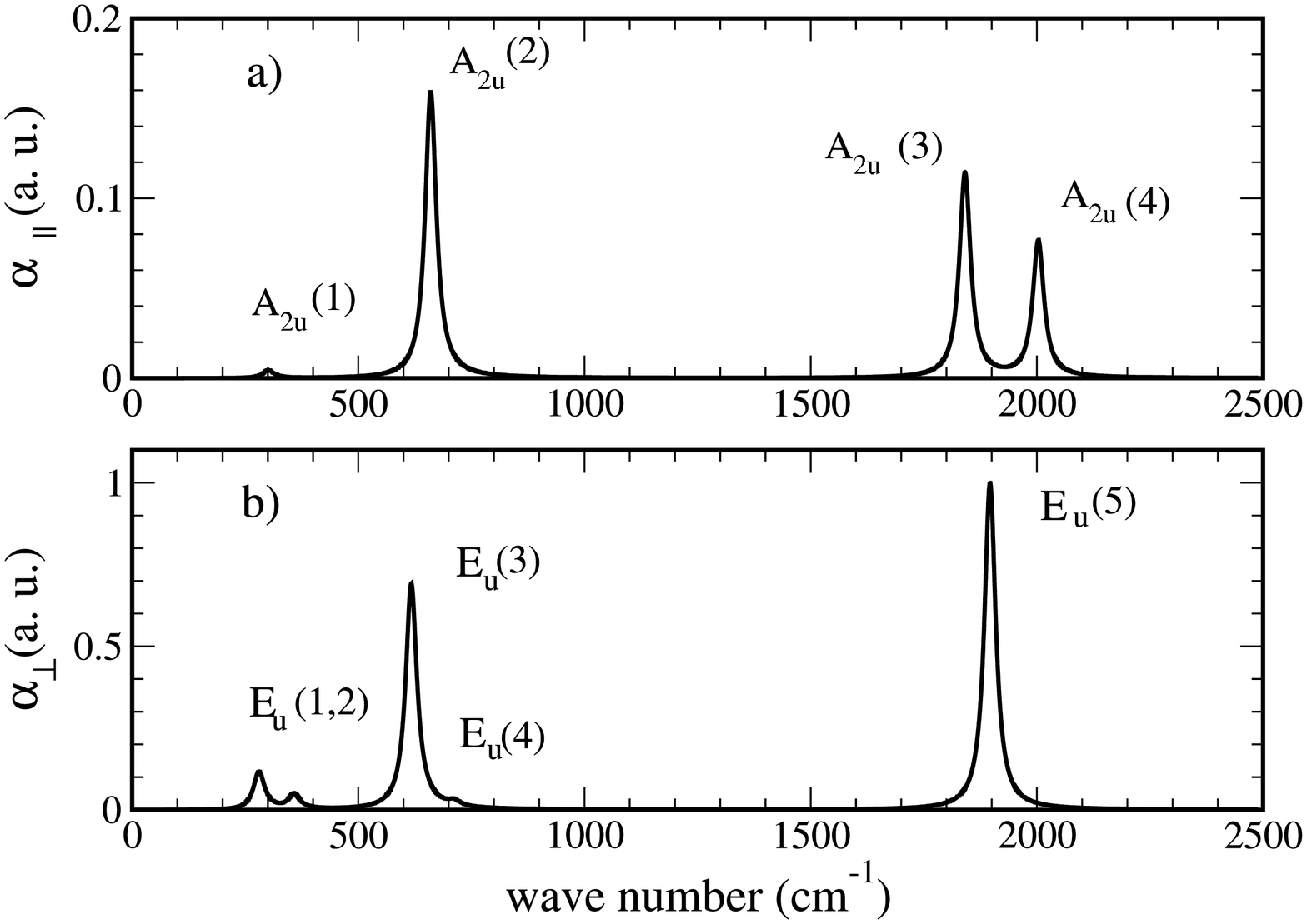}}                                                   
\caption{IR absoption spectrum for a) electric field  {\bf E} $\parallel$ {\bf c} ($\alpha_{\parallel}$) and for
b) {\bf E} $\perp$ {\bf c}  ($\alpha_{\perp}$).  
The Lorenztian broadening is $\gamma$= 15 cm$^{-1}$ (Eq. \ref{lorenz}).}
 \label{abs}                                                                                                
 \end{figure}

The functions  $\alpha_{\perp}(\omega)$ and  $\alpha_{\parallel}(\omega)$ are shown in Fig. \ref{abs}.
The phonons around 700-800 cm$^{-1}$ are bending modes of the Al-H bonds. The highest frequency modes in the range 1800-2000 cm$^{-1}$
are stretching modes of the Al-H bonds. In particular the A$_{2u}$(3) mode is a 
stretching of the Al-H$_1$ apical bond. The other modes
are stretching of the Al-H$_2$ bridging bonds. The displacement pattern of the IR and Raman active modes are shown
in Fig. \ref{disp}. The modes below  360  cm$^{-1}$ (cfr. Table \ref{phonon}) are lattice modes involving a rigid motion of the
teatrahedra.

\vspace{1.cm}

Since the experimental data are available only for a polycrystalline sample 
\cite{ficth2} an angle-averaged absorption coefficient
is needed to compared  theoretical and experimental data.
For a generic {\bf q} forming an angle $\theta$ with the optical axis 
the absorption coefficient for the extraordinary wave can be obtained from the imaginary part of $\epsilon_{\theta}$ 
(Eq. \ref{extra}). For non-polarized light the total absorption coeffient can be equivalently expressed as

\begin{equation} 
\alpha_{\theta}(\omega) =   \frac{2 \pi^2}{V_onc}\sum_{j=1}^{\nu + \mu} | 
		  \sum_{\kappa =1}^N  {\bf \hat{q}} \wedge \tens{Z} \cdot \frac{{\bf \tilde{e}}(j,\kappa)}{\sqrt{M_{\kappa}}}|^2 
		  \delta(\omega-\tilde{\omega}_j),
\label{abstheta} 
\end{equation}

where ${\bf \tilde{e}}(j,\kappa)$ and $\tilde{\omega}_j$ are eigenstates and eigenvalues of the full dynamical matrix
including the non-analytic term (Eq. \ref{macro}) which mixes  E$_u$ and A$_{2u}$ modes.

\begin{figure}[!h]
\centerline{\epsfysize= 5  truecm\epsffile{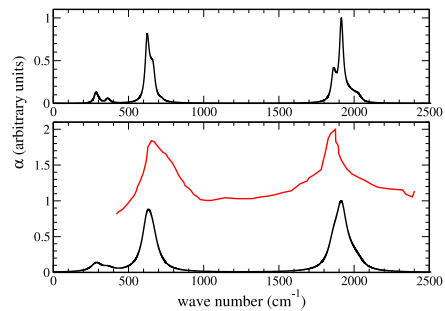}}
\caption{(Color online) IR absorption spectrum for polycrystalline Mg(AlH$_4$)$_2$. 
a) Theoretical spectrum with Lorentzian broadening $\gamma$= 15 cm$^{-1}$ (Eq. \protect\ref{lorenz}).
b) Theoretical spectrum (black line) with $\gamma$=  40 cm$^{-1}$ 
 compared with the experimental spectrum (gray line (red online), adapted from Ref. \protect\cite{ficth1}).
 The experimental spectrum is normalized by setting equal the experimental and theoretical intensity of the
 highest peak around 1800 cm$^{-1}$.}
 \label{absexp}
  \end{figure}

The angle-averaged absorption coefficient
 has been  obtained from Eq. \ref{abstheta} as

\begin{equation}
\alpha_{ave}= \sum_n sin(\theta_n)\alpha_{\theta_n},
\label{absave}                                                                                             
\end{equation}  

where the sum runs over ten angles equally spaced in the range 0-$\pi$.
The resulting theoretical absorption coefficient for a polycrystalline sample is compared with the
experimental IR spectrum in Fig. \ref{absexp}\\

\vspace{1.cm}

Good agreement with experimental data is obtained with a Lorentzian broadening of $\gamma$= 40 cm$^{-1}$.
(Fig. \ref{absexp}), but for the experimental shoulder around 800 cm$^{-1}$ which is absent in the theoretical spectra.
This misfit might be either due to an underestimation of the A$_{2u}$(2) mode which is very close to the strongest
IR mode E$_{2u}$(3) (cf. Fig. \ref{abs}) or to a large inhomogeneous broadening of the experimental spectra.
In fact, the line-width of the experimental peak are very large and might be partially due to residual solvent adducts
which are released only after dehydrogenation \cite{ficth2}.

\subsubsection{Raman spectrum}

The differential cross section  for Raman scattering (Stokes) in non-resonant conditions
is given by the following expression \cite{cardona,bruesh} (for a unit volume of scattering sample)

\begin{equation}
\frac{d \sigma}{d \Omega d \omega} = \sum_{j} \frac{\omega_S^4 }{c^4} | {\bf e_S} \cdot \tens{R}^j \cdot  {\bf e_L}|^2
 (n(\omega)+1)\delta(\omega-\omega_j),
\label{raman}
\end{equation}

where  $n(\omega)$ is the Bose factor,
$\omega_S$ is the frequency of 
the scattered light, ${\bf e_S}$ and ${\bf e_L}$ are the polarization vectors of the scattered and incident
light,
respectively. The  Raman tensor $\tens{R}^j$ associated with the $j$-th phonon is given by

\begin{equation}
R_{\alpha,\beta}^j =  \sqrt{\frac{V_o \hbar}{2 \omega_j}}
\sum_{\kappa=1 }^N \frac{\partial \chi_{\alpha,\beta}^{\infty}}{\partial {\bf r}(\kappa)}
\cdot \frac{{\bf e}(j,\kappa)}{\sqrt{M_{\kappa}}},
\label{ramanT}
\end{equation}

where $V_o$ is the unit cell volume (274.25 $\rm\AA^3$) \cite{ficth3},  ${\bf r}(\kappa)$ is the position of atom $\kappa$-th and 
$\tens{\chi}^{\infty}=(\tens{\varepsilon}^{\infty}-{\bf \delta})/4\pi$
is the electronic susceptibility.
The inner longitudinal macroscopic electric field has no effect on $g$-modes. 
The tensor $\tens{R}^j$ is computed from $\tens{\chi}^{\infty}$ by finite differences, by moving the atoms along the phonon
displacement pattern with maximum displacement of 0.002 $\rm\AA$.
The Raman tensor (Eq. \ref{raman}) for the Raman-active  irreducible representations has the form

\begin{displaymath}
A_{1g} \Rightarrow
\left [ 
\begin{array}{ccc}
a  & . & . \\
.  & a & . \\
.  & . & b \\
\end{array}
\right ]
\label{a1g}
\end{displaymath}

\begin{displaymath}
E_g, 1 \Rightarrow
\left [ 
\begin{array}{ccc}
 c  & .              & .   \\
.              &  - c  & d   \\
.              & d              & . \\
\end{array}                                                                                          
\right ]                                                                                             
\label{eg1}
\end{displaymath}

\begin{displaymath}
E_g, 2 \Rightarrow                                                                                   
\left [                                                                                              
\begin{array}{ccc}                                                                                   
. & -c              & -d   \\                                                             
-c              & . & .   \\                                                             
-d              & .              & . \\                                                             
\end{array}                                                                                          
\right ]                                                                                             
\label{eg2}
\end{displaymath}

The coefficients $a$,$b$,$c$ and $d$ calculated from first principles as outlined above are given for each mode in Table I.

The experimental Raman spectrum \cite{ficth2} is available only for non-polarized light and backscattering geometry 
on a polycristalline sample.
To compare with experimental data, Eq. \ref{raman} must be integrated over the solid angle by summing over
all possible polarization vectors  ${\bf e_S}$ and ${\bf e_L}$ consistent with the backscattering geometry.
The  total cross section for unpolarized light in backscattering geometry is obtained from Eq. \ref{raman}
with the substitution

\begin{eqnarray}
4 &(&R_{xx}^2+R_{yy}^2+R_{zz}^2) +7(R_{xy}^2+R_{xz}^2+R_{yz}^2)+ \nonumber \\
&(&R_{xx}R_{yy}+R_{xx}R_{zz}+R_{zz}R_{yy})
\rightarrow  | {\bf e_S} \cdot \tens{R}^j \cdot  {\bf e_L}|^2 \nonumber\\
\end{eqnarray}

\begin{figure}[!h]
\centerline{\epsfysize= 5. truecm\epsffile{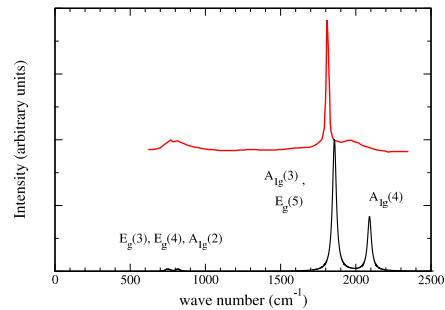}}                             
\caption{(Color online) Theoretical (black line) and experimental (gray line (red online), adapted from Ref. \protect\cite{ficth2}) Raman spectra of
 polycristalline Mg(AlH$_4$)$_2$ for unpolarized light in backscattering geometry.
 The experimental spectrum is normalized by setting equal the experimental and theoretical intensity of the
 highest peak around 1800 cm$^{-1}$.}
 \label{ramanexp}   
  \end{figure}

The resulting theoretical Raman spectrum is compared with experimental data in Fig. \ref{ramanexp}.

\vspace{1.cm}

The agreement is good, but for the highest frequency peak which is too high in energy and in intensity with
respect to experiments. The displacement pattern of the Raman active intra-tetrahedra modes are sketched in Fig. \ref{disp}.
The strongest Raman peak is due to the A$_{1g}$ stretching mode of the Al-H$_2$ bond at 1845 cm$^{-1}$.

\section{Conclusions}

Based on Density Functional Theory we have optimized the structure of Mg(AlH$_4$)$_2$. Due to the lack of 
van der Waals forces within the current approximations to the energy functional, the interlayer spacing between the
neutral AlH$_4$-Mg-AlH$_4$ sheets stacked along the $c$ axis is largely  overestimated (6 $\%$) in our calculations.
Conversely, by fixing the lattice parameters to the experimental ones, the optimization of the internal structure
provides a geometry in fair agreement  with that inferred experimentally from x-ray powder diffraction data. The main 
discrepancy regards the position of the hydrogen atoms (which however can not be detectd accurately from x-ray powder
diffraction) resulting in Al-H bond lengths
which differ up to 0.1 $\rm\AA$ from the exprimental ones.
As a consequence,  the AlH$_4$ tetrahedra are much more symmetric in the theoretical geometry than in that proposed
experimentally. However, the IR and Raman spectra calculated within density functional perturbation theory are in
good agreement with the experimental spectra which supports the correctness of the crystal structure emerged from the
ab-initio calculations.

\section{Aknowledgments}

We gratefully thank G. Benedek, V. Boffa, G. Dai, V. Formaggio and S. Serra for discussion and information.
This work is partially supported by  the INFM Parallel Computing Initiative.


\begin{thebibliography}{169}
\bibitem{notadebye}  The zero-point energy per unit cell of the acoustic bands within the Debye model is given by
$E_o=9/8k_B\Theta_D= (3^7/2^{11}\pi)^{1/3} h  \bar{c}  V_o^{-1/3}$ (Ref. \protect\cite{ash}), 
where $\Theta_D$ is the Debye temperature, $V_o$ the unit cell volume and $\bar{c}$ is the phase velocity of the
acoustic branches (1/3 $v_{LA}$ + 2/3 $v_{TA}$) averaged over all directions.
For Al $\Theta_D$= 394 K. For MgH$_2$ $\bar{c} \approx$ 4.5 km/s, from the phonons calculation in Ref. \protect\cite{phonMgh2}
(by assuming isotropic phase velocities). For Mg(AlH$_4$)$_2$,  $\bar{c} \approx$ $v_{Z}/3$ + $v_{L}2/3$, where 
 $v_{Z}$=  3.75 km/s and $v_{L}$= 4.38 km/s are the phase velocities (averaged over the TA and LA branches) along the
 $\Gamma$Z and $\Gamma$L directions of the BZ, calculated within dendisty functional pertirbation theory.
 The zero point energy per unit cell of the optical modes of Mg(AlH$_4$)$_2$, MgH$_2$ and H$_2$ are 
 162.4, 73.2, and 26.2 kJ/mol, respectively.
The contribution of the vibrational modes to the free energy at finite temperature has been
computed by considering only the optical phonons at the $\Gamma$-point and the acoustic phonons within the Debye approximation as
given above.
\end{thebibliography}
\end{document}